%% file: jem-euso-skeleton-bibtex.tex
\documentclass[a4paper,11pt]{article}
\usepackage{pos}

\usepackage{natbib}
\title{The EUSO-TA ground-based detector: results and perspectives}
 \ShortTitle{The EUSO-TA}

\author*[a,b]{Zbigniew Plebaniak}

\affiliation[a]{Universit\`{a} di Roma Tor Vergata - Dipartimento di Fisica\\
  Via della Ricerca Scientifica 1 – 00133, Roma, Italy}

\affiliation[b]{Istituto Nazionale di Fisica Nucleare - Sezione di Roma Tor Vergata\\
Via della Ricerca Scientifica 1 – 00133, Roma, Italy}

\onbehalf{for the JEM-EUSO Collaboration\\[-1mm]{\normalsize \normalfont (a complete list of authors can be found at the end of the proceedings)}}

\emailAdd{zbigniew.plebaniak@roma2.infn.it}

\abstract{
EUSO--TA is a ground-based telescope installed in 2013 in the Black Rock Mesa Telescope Array (BRM-TA) site, operating with 2.5~$\mu$s time resolution to observe the night sky in the UV range. 
The optical system contains two 1~m$^2$ Fresnel lenses providing to the telescope a field of view of $11^\circ \times 11^\circ$. 
Signals are focused on the Photo Detector Module (PDM), with the focal surface composed of 36 Hamamatsu Multi-Anode PhotoMultiplier Tubes (MAPMTs), with 64 pixels/anodes each.
The telescope is housed in a shed in front of the BRM-TA fluorescence detectors, and it is viewing towards azimuth $\sim307^\circ$.
The main aim of the experiment is to validate the design of the JEM-EUSO detectors and firmware with the final goal of observing ultra-high-energy cosmic rays (UHECRs) from space.
Since the first installation of the EUSO-TA detector, 9 UHECR events have been detected and confirmed by comparison with TA observations.
The night-sky UV background in different conditions, signals from stars and meteors have been measured, and anthropogenic signals, such as calibration lasers or planes.
In 2019 an upgrade of the detector to a EUSO-TA2 version began, with a Covid brake till 2022.
The new configuration will allow for more frequent and specialized observations. 
In this work, we present the status and perspectives of the EUSO-TA experiment, including a discussion of recently obtained results.
}

\ConferenceLogo{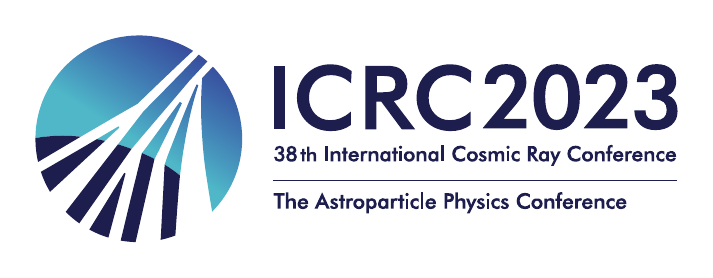}

\FullConference{%
38th International Cosmic Ray Conference (ICRC2023)\\
  26 July - 3 August, 2023\\
  Nagoya, Japan}

\begin{document}
\maketitle

\section{Introduction}

\begin{figure}[ht]
\centering
\includegraphics[width=0.495\textwidth]{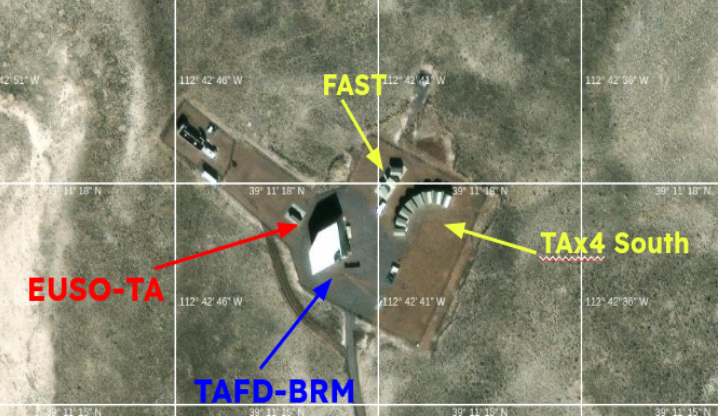}
\hspace{0.12cm}
\includegraphics[width=0.483\textwidth]{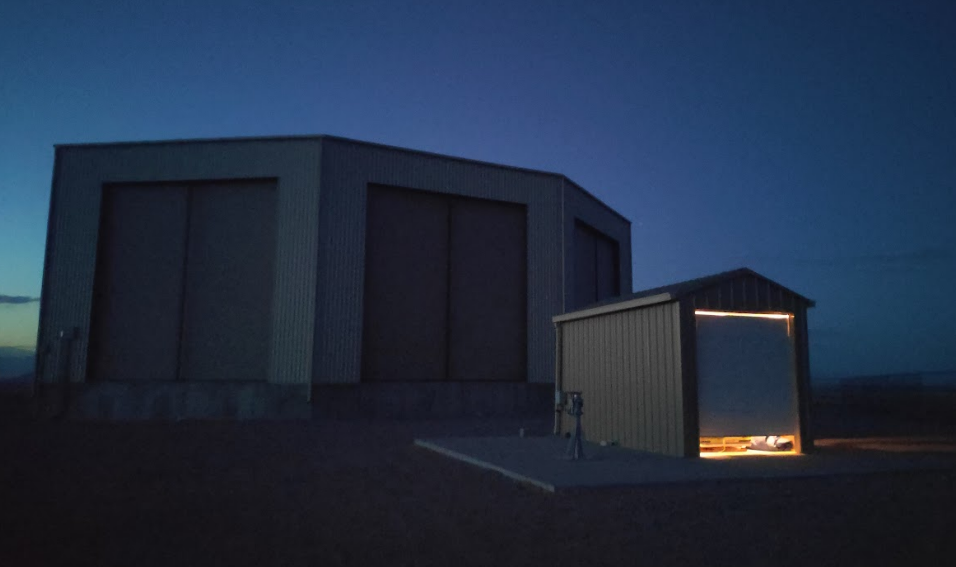}\\

\caption{Location of EUSO-TA experiment on the Telescope Array Black Rock Mesa site. The position of the EUSO-TA dome is marked on the left plot concerning the TA installations and FAST experiment. On the right plot, the EUSO-TA dome is present against the background of the BRM-TAFD station.}
\label{EusoTaLocation}
\end{figure}

EUSO-TA is a small ground-based fluorescence telescope installed in the Black Rock Mesa Telescope Array \cite{TAMEDA200974} to observe the night sky in the UV band in parallel with Telescope Array Fluorescence Detector 
(TA--FD).
The location is shown in the Figure~\ref{EusoTaLocation}.
The experiment is a part of the JEM-EUSO 
(Joint Exploratory Missions for 
an  
Extreme Universe Space Observatory) scientific program~\cite{JEM-EUSO-program} whose primary goal is to detect Ultra High Energy Cosmic Rays (UHECRs) from space by detecting UV emission from Extensive Air Showers (EAS).
EUSO-TA
optical system and mechanical structure were installed in March 2013, while the first detector was mounted and run in February 2015, 
detecting the first calibration sources. 
During its operation in 2015 - 2016, six observation campaigns took place. 
The first UHECR events were registered as triggered by the 
(TA--FD)
station in May 2015.
In October 2015 EUSO-TA telescope was operating with Photo Detector Module (PDM) used previously in the EUSO-Balloon experiment~\cite{eusobal}.
During measurement sessions, artificial light sources like Telescope Array Central Laser Facility (CLF) or Global Laser System (GLS) - a mobile UV laser provided by Colorado School of Mines~\cite{Hunt:2015rzj} have been used for calibration as well as many stars in the field of view. 
The EUSO-TA telescope upgrade to TA2 started in 2019, but the installation at the TA--FD site was completed in 2022, since COVID restrictions.

\subsection{EUSO-TA telescope}
The EUSO-TA optical system with an overall efficiency of about 40$\%$ and a field of view of 10.6$^\circ$ x 10.6$^\circ$ is composed of two Fresnel lenses (with 0.92 m$^2$ of active area) fabricated from UV transmitting polymethyl-methacrylate (PMMA).
The lenses focus light onto the PDM focal surface composed of a matrix of 36 Multi-Anode Photomultiplier Tubes (MAPMTs), each containing 64 channels manufactured by Hamamatsu. 
A UV-transmitting band pass filter (Schott BG3) covers the photomultiplier limiting observed wavelengths to 290 - 430 nm.
PDM, with 2304 pixels, a 2.88 mm side each, is mounted at about 43 cm from the second lens.
The field of view of a single pixel is 0.2$^\circ$ x 0.2$^\circ$. 
The EUSO-TA detector operated in single photon counting mode with a double pulse resolution of about 30ns. The PMTs were read by SPACIROC1 Application-Specific Integrated Circuits (ASIC)~\cite{Miyamoto:2013cxc}. One frame contains counts summarized over Gate Time Unit (GTU) of 2.5 $\mu s$, including 200 ns of a dead time. The elevation angle of the telescope can be manually changed between 0$^\circ$ and 25$^\circ$.

\section{EUSO-TA results}
The EUSO-TA detector operated in two trigger modes: self-triggered with a 1Hz trigger rate or using an external trigger from the TAFD-BRM station.
Based on the taken data, an internal UHECR dedicated first-level trigger has been tested as well~\cite{Abdellaoui:2017huh}. 
In case of receiving the trigger, EUSO-TA saved one packet containing 128 GTU frames around the triggering moment from buffer.
An average trigger rate from TA was about 4 Hz.
As the saturation level for the EUSO-TA detector is about 28 counts/GTU, single and working properly pixel counts are distributed according to the Poisson Law~\cite{Abdellaoui:2018rkw}.

\subsection{Detection of CRs}
\begin{figure}[ht]
\centering
\includegraphics[width=1.01\textwidth]{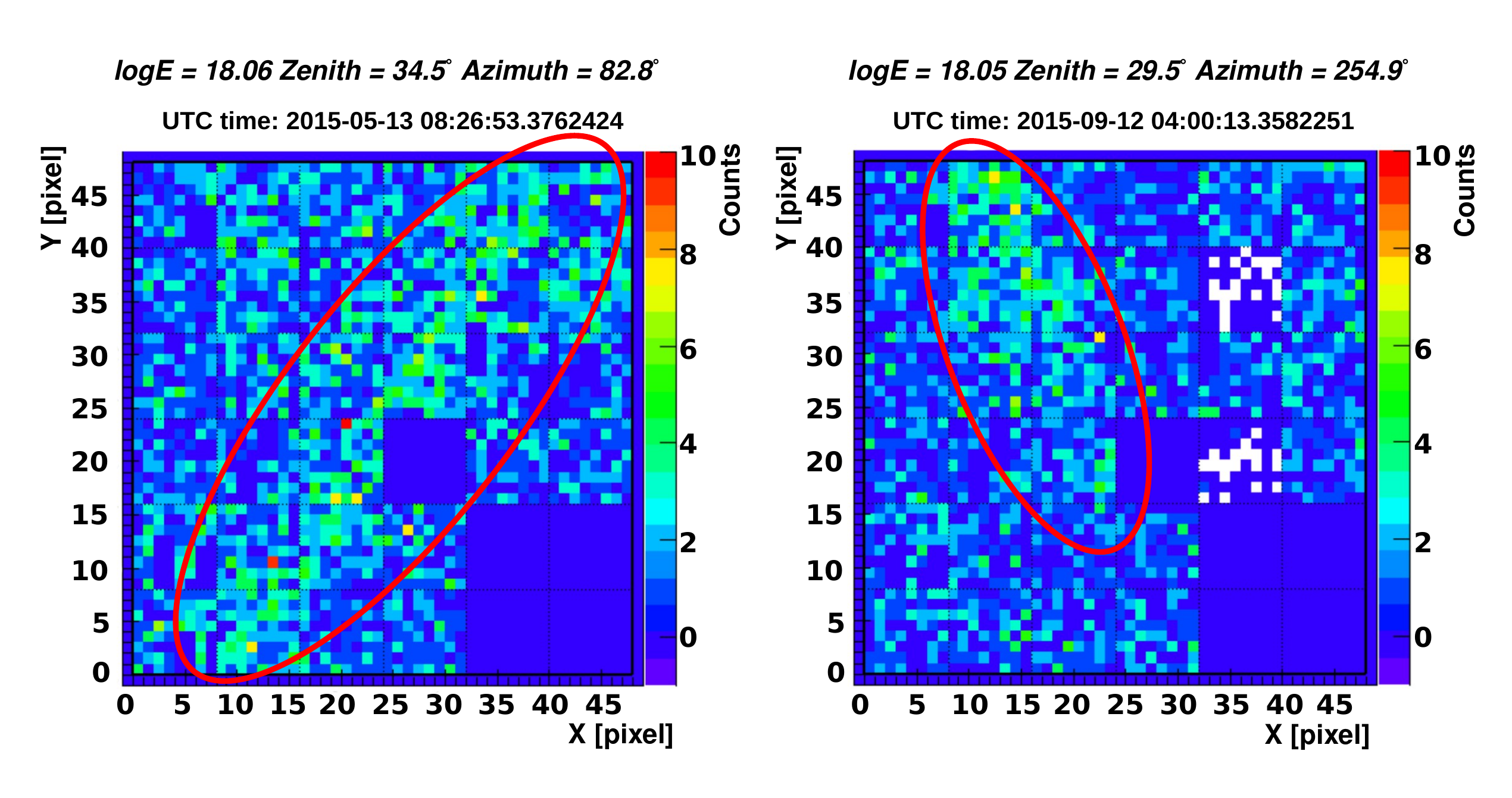}\\

\caption{Two of the nine UHECR observed by EUSO-TA with the same energy of about 10$^{18}$ eV but different orientations regarding the detector. Red ellipses mark regions with recognized signals. In both cases, the signal has been detected in one frame only. In the presented plots, only a part of the EAS signal is visible due to the limited field of view of the EUSO-TA detector. }
\label{DetectedCrEvents}
\end{figure}

During four EUSO-TA campaigns in 2015-2016, covering 140 hours of observation, nine UHECRs have been found in data triggered externally. 
TA has confirmed all events.
The energy of registered events varies from 10$^{17.7}$ to 10$^{18.8}$ eV for the distance from the telescope between 1 km and 9 km.   
In Figure~\ref{DetectedCrEvents}, we show as an example two of them. 
The event shown in the left panel, detected on May 13th 2015, is the first event seen and recognized as UHECR using the JEM-EUSO technology.
Most events are visible in the single frame due to their short distance from the detector.
In the case of one event only, the signal is clearly visible in two frames.
More detailed analysis of detected events can be found in~\cite{JEM-EUSO:2019gvd}.
EUSO-TA usually observes a very late stage of the EAS development.
All detected events have been studied with ESAF and OffLine simulations and are crucial in designing future EUSO missions and developing the trigger algorithms. 

\subsection{Other events, calibration with lasers and stars}
\begin{figure}[ht]
\centering
\includegraphics[width=0.35\textwidth]{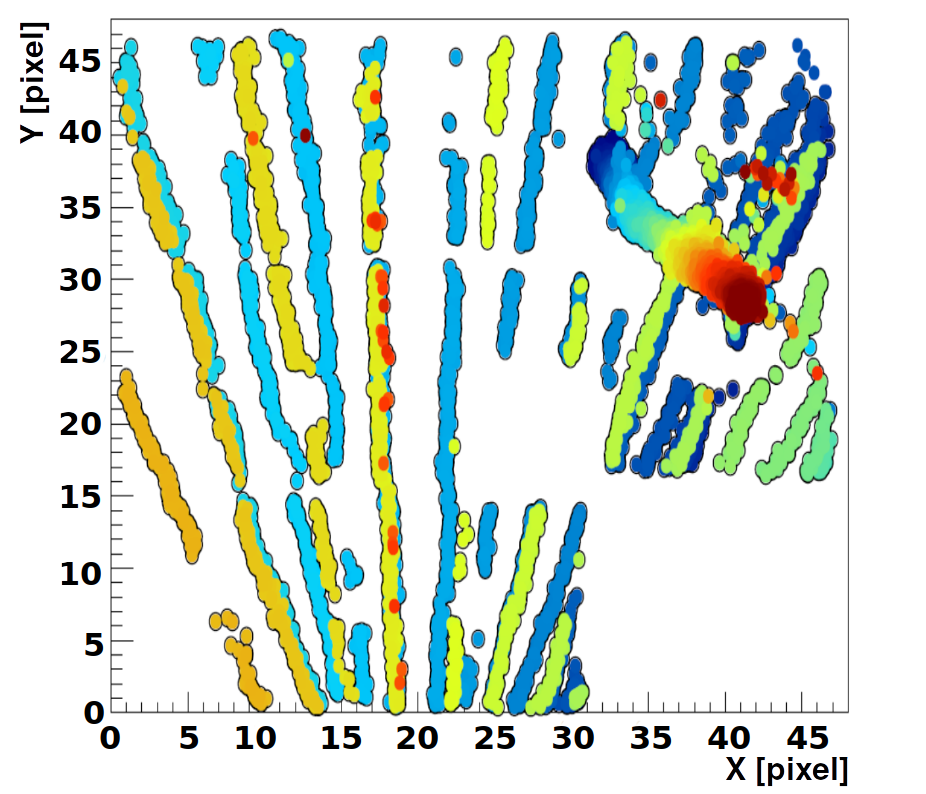}
\includegraphics[width=0.64\textwidth]{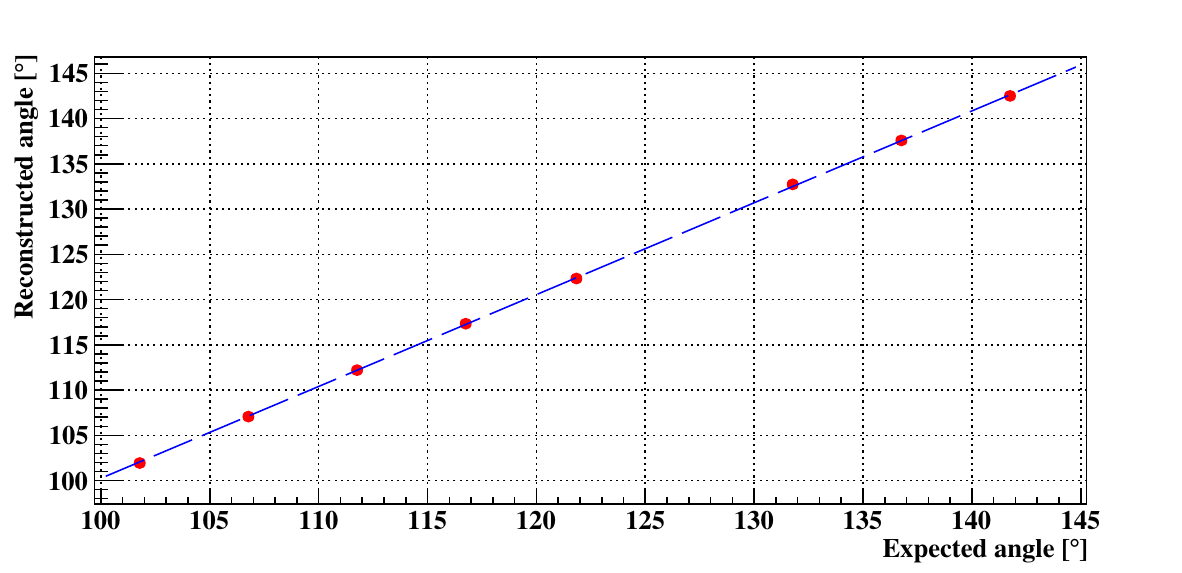}\\

\caption{Left: Positions of GLS signals detected by the offline trigger while swapping the EUSO-TA focal surface. Colours indicate the firing time (blue: early, red: late). In the central part, a few red points indicate triggers from CLF firing during the GLS session. Additional triggers in the top right corner come from an aeroplane flying in the field of view. Right: accuracy of angular reconstruction of GLS shots.}
\label{EusoTaGlsReco}
\end{figure}

Observing the night sky to search for UHECR signals, EUSO-TA detected various natural signals like lightning, meteors and stars and anthropogenic ones like planes in the field of view, calibration diodes and lasers.
Five meteors in the field of view with a limited magnitude of 5.5$^m$ have been detected by chance as the TAFD detector does not trigger them.
Laser signals produced by CLF and GLS mobile systems have been registered similarly. 
During TA operations, CLF shoots vertically every half an hour for 30s at 10 Hz, generating UV pulses with an energy of 4-6 mJ (2.2~mJ corresponds to a 10$^{19}$~eV shower seen from 21 km).
GLS laser provided by the Colorado School of Mines worked in 2 - 22 mJ energy ranges.
Registered laser signals have been recognized in EUSO-TA data using an offline trigger algorithm (see Figure~\ref{EusoTaGlsReco}). 
The EUSO-TA detector indicates the linear dependency of the registered signal on the intensity of the GLS beam in the range of 4 - 22 mJ.

\begin{figure}[ht]
\centering
\includegraphics[width=0.60\textwidth]{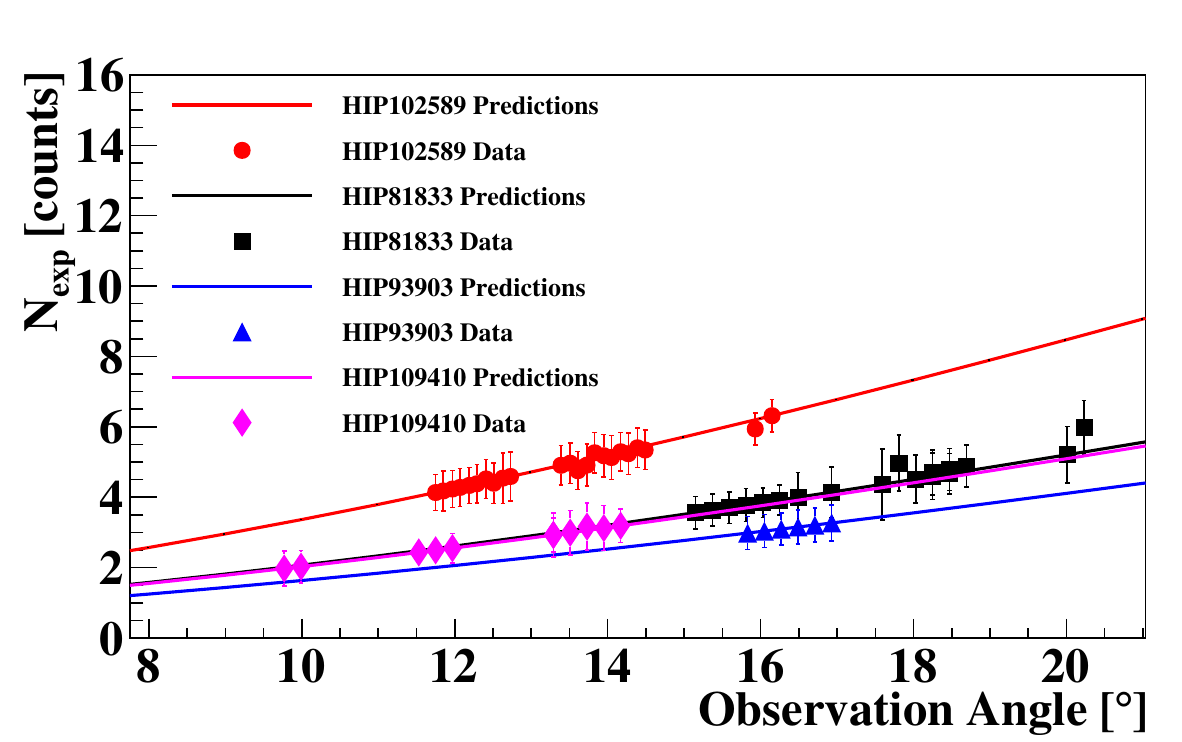}
\includegraphics[width=0.39\textwidth]{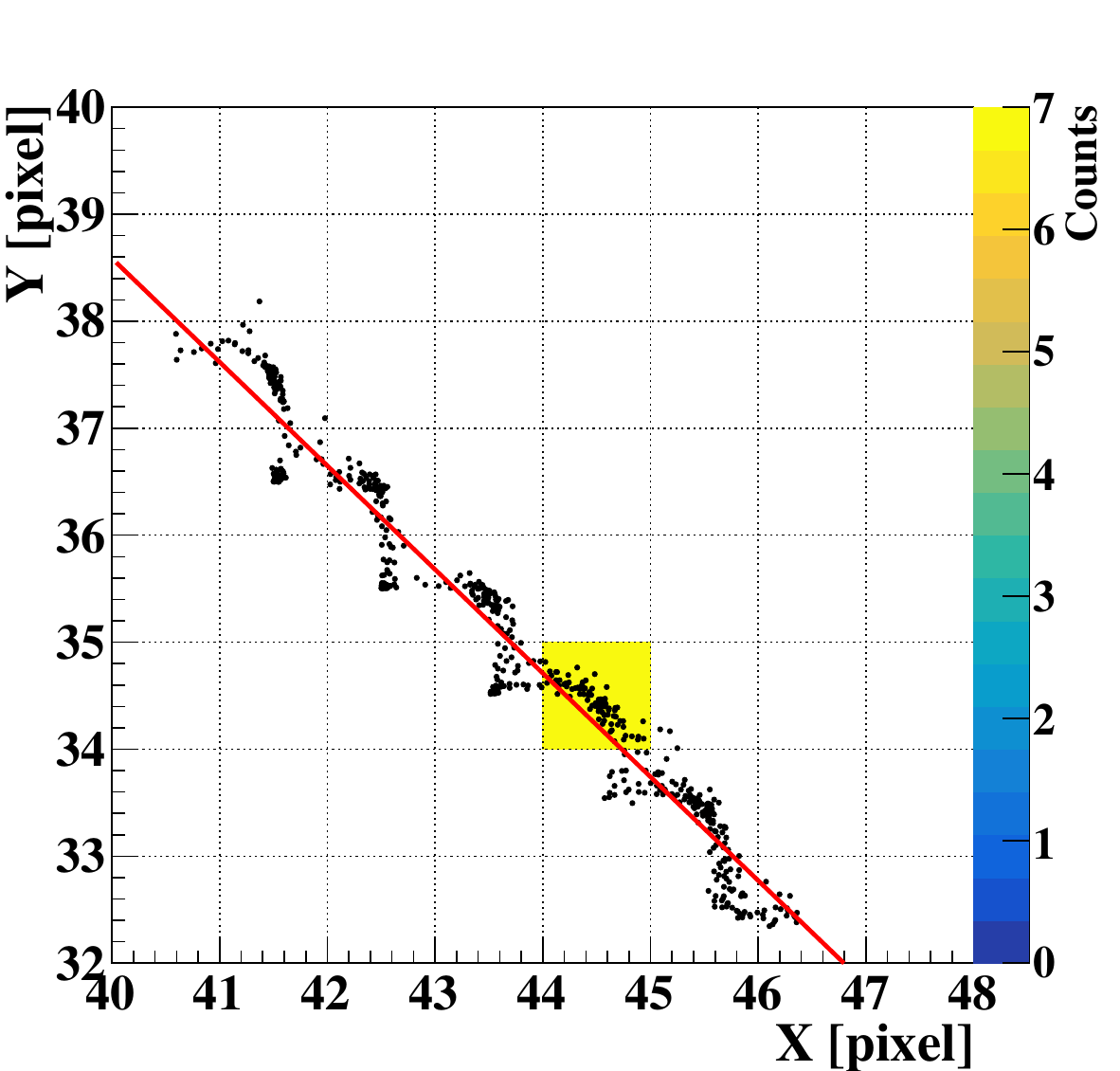}\\

\caption{Left: comparison between expected (lines) and measured (and calibrated, points) signal for four stars in the field of view during the EUSO-TA campaign in October 2015. Predictions have been calculated using the Pickles stellar spectral flux library, and atmospheric transmission calculated with libRadTran package Right: track of the bright star with apparent magnitude 2.23$^{m}$ fitted on the focal surface. The position of the signal during transition can be determined with accuracy better than 0.4 pixels.}
\label{StarAnalysis}
\end{figure}

During night measurements, EUSO-TA observed many stars in the field of view as point-like sources giving strong signals in the UV band.
Stars with known spectra have been used as calibration sources.
Stars can also be used for probing the focal surface during operations.
In Figure~\ref{StarAnalysis}, we compare the measured and expected signal and the star track fitted on the focal surface.
In the analyzed part of the data, an average registered background of the night sky ranged from 1.09 to 1.91 counts/pixel/GTU. 
The average detector efficiency has been calculated as (4.10$\pm$0.49)$\%$ in the range 300nm - 400nm~\cite{stars-icrc2023}.


\section{Common operations with other EUSO experiments}
{
EUSO-TA installation is essential for the JEM-EUSO collaboration since it provides an opportunity to observe UHECRs and to test other EUSO detectors.
The infrastructure allows integration, reconfiguration and fixing of EUSO detectors when necessary.
Until now, EUSO-TA was operating with its original PDM and the one used in the EUSO-Balloon experiment~\cite{eusobal}.
In September 2016, common campaign with EUSO-SPB1 experiment took place (see Fig.~\ref{EusoSpb1AndTa})~\cite{Adams:2020vtp}.
During the eight-night campaign, EUSO-SPB1 was prepared for observations and tested with calibration sources and stars from the night sky to optimize the detector parameters before the final mission.
During this time, EUSO-TA was refurbished, prepared to operate and finally observed with EUSO-SPB1 in the campaign's last days.
In August and September 2022, a similar operation was carried out with the EUSO-SPB2 detector containing 3 PMDs~\cite{euso-spb2-field-icrc2023}.

\begin{figure}[ht]
\centering
\includegraphics[width=0.89\textwidth]{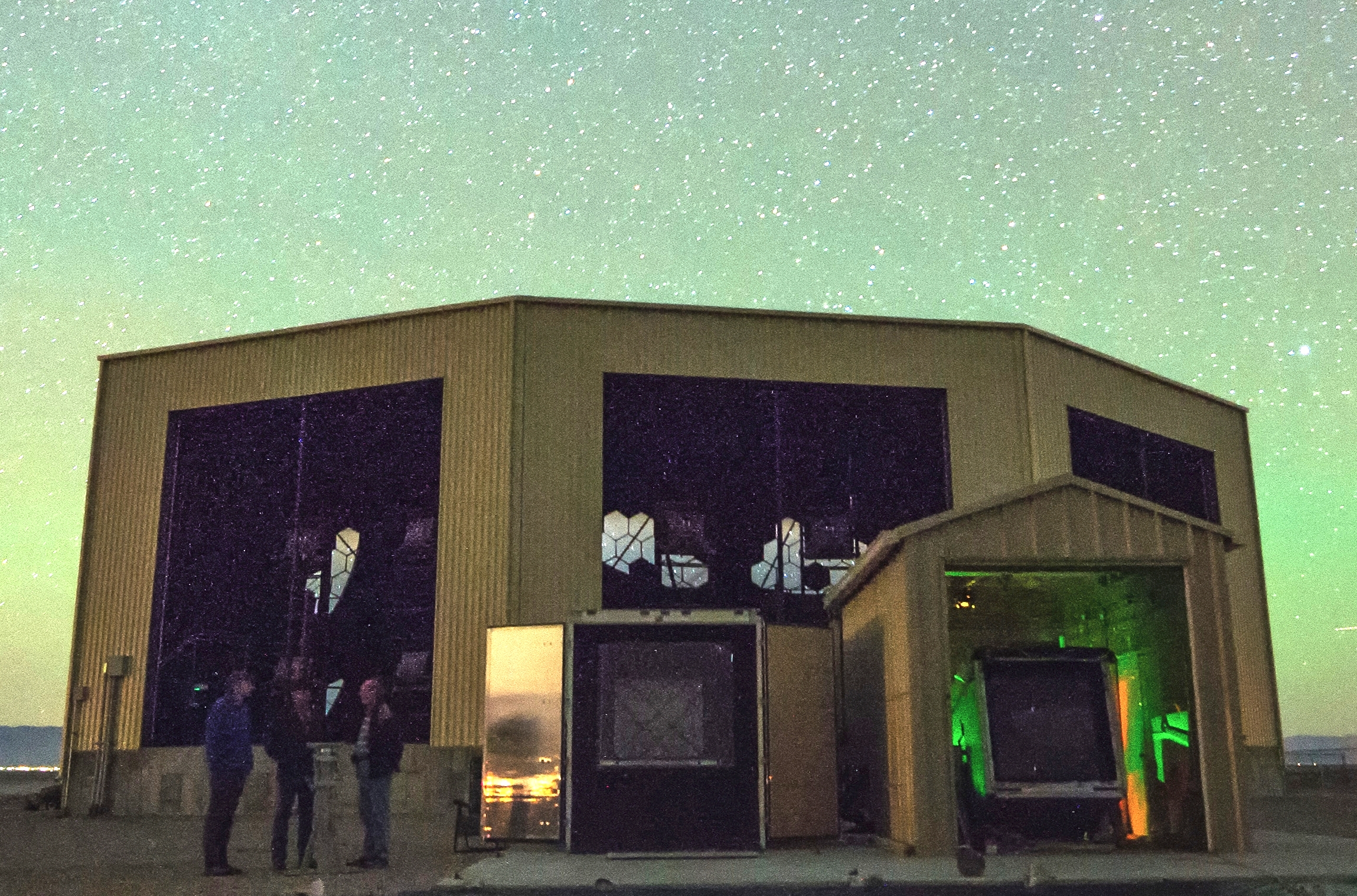}
\caption{EUSO-TA telescope (right) next to EUSO-SPB1 (left) detector in front of TAFD-BRM station. Photography by M. Mustafa was taken during a cross-calibration field test in October 2016 when all three detectors were observing the night sky. }
\label{EusoSpb1AndTa}
\end{figure}

\section{Upgrade to EUSO-TA2}
In June 2022, an upgraded version of the EUSO-TA detector was installed on the TA site.
Conserving the old version of the optical and mechanical structure, the PDM has been changed to the new one with improved parameters.
Fig. \ref{EusoTa2PdmUpgrade} shows the new PDM during tests and installation.
We used elementary cells with MAPMTs used previously in the EUSO-Balloon experiment.
However, ASICs have been upgraded to SPACIROC3 with a dead time of 50ns per GTU and 5ns of double pulse resolution~\cite{Blin-Bondil:2014uct}.
SPACIROC3 has been already used in EUSO-SPB1~\cite{Eser:2019ciy} and Mini-EUSO experiments~\cite{Bacholle:2020emk}.
The FPGA board has been upgraded to Xilinx Zynq based. 
The acquisition system and firmware in EUSO-TA2 are identical to the one used in Mini-EUSO and allow for taking data in three temporal resolutions: of 2.5$\mu s$ (D1 GTU), 320$\mu s$ (D2 GTU), and 40.96 ms (D3 GTU).
During the first four campaigns in 2022 and 2023, the EUSO-TA2 detector was set up and tuned.
Almost 50 hours of calibration data have been taken. 
D3 acquisition mode allowed for observing many meteors with a magnitude < 6.8 and stars with M$_{B}$ < 7.3 that can be recognized and analyzed. 
A very preliminary analysis of a few stars in the field of view estimates the detector efficiency of (4.33$\pm$0.87)$\%$.
High uncertainty comes from the low number of analyzed events.
Acquisition in D1 mode in future will allow searching for UHECRs with dedicated trigger algorithms under development.
First estimations of the energy limit for detecting UHECRs by EUSO-TA2 have been set as 10$^{18}$ - 10$^{19}$ eV~\cite{euso-ta2-limits-icrc2023}.

\begin{figure}[ht!]
\centering
\includegraphics[width=0.48\textwidth]{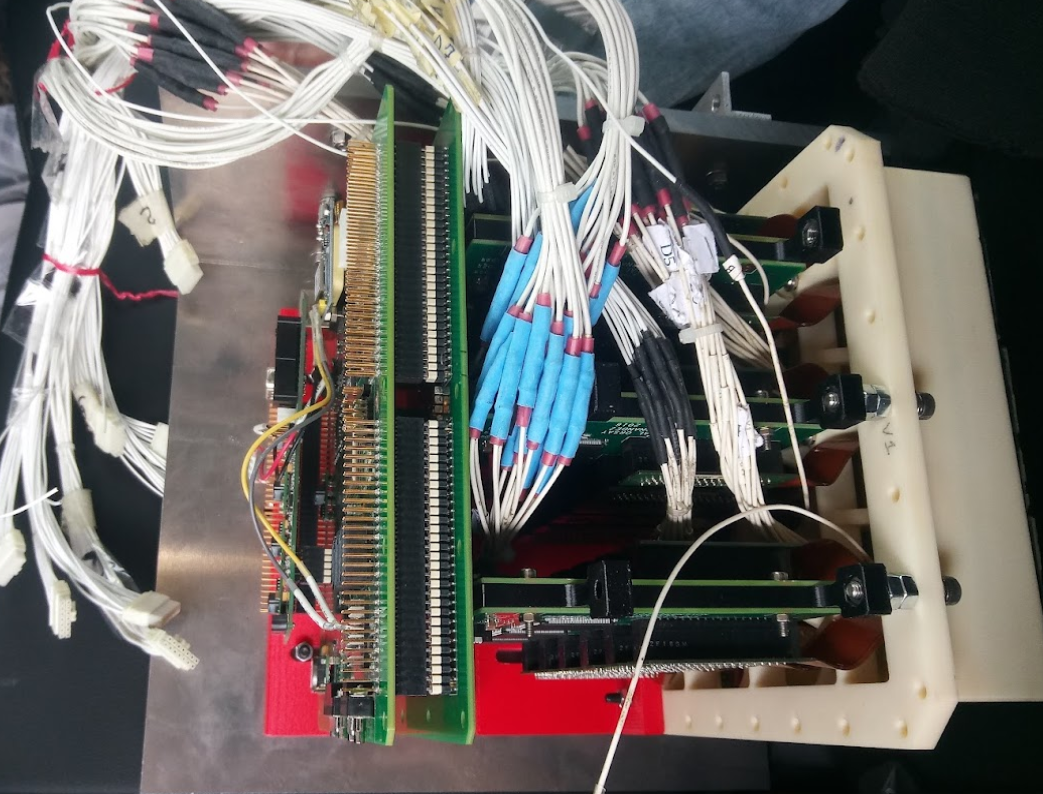}
\hspace{0.25cm}
\includegraphics[width=0.49\textwidth]{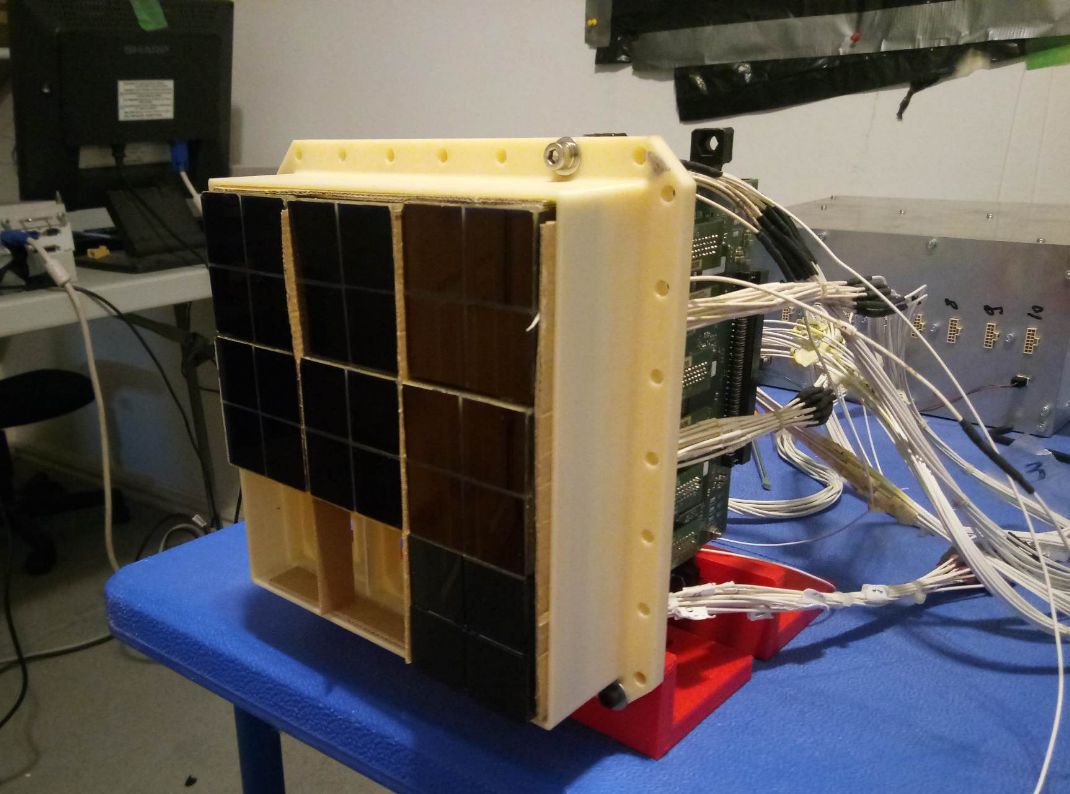}\\

\caption{EUSO-TA2 detector during operations before mounting in the telescope. The left picture shows integrated PDM with visible PDM board and EC boards. Photomultipliers on the focal surfaces are mounted in white plastic cover. The right plot shows PDM while replacing two unstable ECs with a silver HVPS module in the background.}
\label{EusoTa2PdmUpgrade}
\end{figure}

{
\color{red}
}

\section{Conclusions}
For the years, EUSO-TA played an essential role in the JEM-EUSO program allowing for testing of the EUSO technology.
With the first version of EUSO-TA, we confirmed that using Fresnel lenses with a focal surface composed of MAPMTs works to observe fast-moving EAS events in the atmosphere. 
Detected UHECR events, stars, meteors, and calibration sources gave the first estimations of the detector sensitivity for measurements in atmospheric conditions. 
An upgraded EUSO-TA2 will allow for testing more advanced solutions in triggering UHECR events and hardware development.


\small{\bf Acknowledgments:}
This work was partially supported by Basic Science Interdisciplinary Research
Projects of RIKEN and JSPS KAKENHI Grant (JP17H02905, JP16H02426 and
JP16H16737), by the Italian Ministry of Foreign Affairs and International Cooperation, 
by the Italian Space Agency through the ASI INFN agreements Mini-EUSO n. 2016-1-U.0, EUSO-SPB1 n. 2017-8-H.0, OBP (n. 2020-26-Hh.0), EUSO-SPB2 n. 2021-8-HH.0 and by ASI INAF agreement n. 2017-14-H.O, by NASA awards and grants 11-APRA-0058, 16-APROBES16-0023, 17-APRA17-0066, NNX17AJ82G, NNX13AH54G, 80NSSC18K0246, 80NSSC18K0473, 80NSSC19K0626, 80\-NSSC\-18K\-0464 and 80NSSC22K1488 in the USA, Deutsches Zentrum f\"ur Luft- und Raumfahrt, by the French space agency
CNES, the Helmholtz Alliance for Astroparticle Physics, funded by the
Initiative and Networking Fund of the Helmholtz Association (Germany), by National Science Centre in Poland grant no 2017/27/B/ST9/02162 and
2020/37/B/ST9/01821. L. W. Piotrowski acknowledges financing by the Polish National Agency for Academic Exchange within Polish Returns Programme no. PPN/PPO/2020/1/00024/U/00001 and National Science Centre, Poland grant no. 2022/45/B/ST2/02889.
The Russian team is supported by ROSCOSMOS, and "KLYPVE" is included into the
A long-term program of Experiments on board the Russian Segment of the ISS.
Sweden is funded by the Olle Engkvist Byggm\"astare Foundation.

\bibliography{my-bib-database}

\input{JEM-EUSO_Authors_July2023.tex}



%
%
%

\end{document}

%% file: JEM-EUSO_Authors_July2023.tex
\newpage
{\Large\bf Full Authors list: The JEM-EUSO Collaboration\\}

\begin{sloppypar}
{\small \noindent
S.~Abe$^{ff}$, 
J.H.~Adams Jr.$^{ld}$, 
D.~Allard$^{cb}$,
P.~Alldredge$^{ld}$,
R.~Aloisio$^{ep}$,
L.~Anchordoqui$^{le}$,
A.~Anzalone$^{ed,eh}$, 
E.~Arnone$^{ek,el}$,
M.~Bagheri$^{lh}$,
B.~Baret$^{cb}$
% ...
}
\end{sloppypar}
\vspace*{.3cm}

{ \footnotesize
\noindent
% ...
$^{cb}$ Universit\'e de Paris, CNRS, AstroParticule et Cosmologie, F-75013 Paris, France\\
% ...
$^{ed}$ Istituto Nazionale di Fisica Nucleare - Sezione di Catania, Italy\\
% ...
$^{eh}$ INAF - Istituto di Astrofisica Spaziale e Fisica Cosmica di Palermo, Italy\\
% ...
$^{ek}$ Istituto Nazionale di Fisica Nucleare - Sezione di Torino, Italy\\
$^{el}$ Dipartimento di Fisica, Universit\`a di Torino, Italy\\
% ...
$^{ep}$ Gran Sasso Science Institute, L'Aquila, Italy\\
% ...
$^{ff}$ Nihon University Chiyoda, Tokyo, Japan\\ 
% ...
$^{ld}$ University of Alabama in Huntsville, Huntsville, AL, USA\\
$^{le}$ Lehman College, City University of New York (CUNY), NY, USA\\
% ...
$^{lh}$ Georgia Institute of Technology, USA\\
% ...
}